\documentclass[onecolumn]{emulateapj}
 \slugcomment{Submitted to ApJ}
\usepackage[hyperfootnotes=false,colorlinks=true,linkcolor=blue,citecolor=blue]{hyperref}
\shortauthors{Thommes, Bryden, Wu \& Rasio} \shorttitle{Breaking mean-motion resonances}

\begin{document}

\title{From mean-motion resonances to scattered planets: Producing the Solar System, eccentric exoplanets and Late Heavy Bombardments}

\author{Edward W. Thommes$^1$, Geoffrey Bryden$^2$, Yanqin Wu$^3$ \& Frederic A. Rasio$^1$ }
\affil{$^1$Department of Physics and Astronomy, Northwestern University, Evanston, IL 60208, USA}
\affil{$^2$Jet Propulsion Laboratory, Pasadena, CA 91109, USA}
\affil{$^3$Department of Astronomy and Astrophysics, University of Toronto, Ontario M5S 3H4, Canada}
\email{thommes@northwestern.edu, Geoffrey\_Bryden@jpl.nasa.gov, wu@astro.utoronto.ca, rasio@northwestern.edu}


\begin{abstract} 
We show that interaction with a gas disk may produce young planetary systems with closely-spaced orbits, stabilized by mean-motion resonances between neighbors.  On longer timescales, after the gas is gone, interaction with a remnant planetesimal disk tends to pull these configurations apart, eventually inducing dynamical instability.    We show that this can lead to a variety of outcomes; some cases resemble the Solar System, while others end up  with high-eccentricity orbits   reminiscent of the observed exoplanets.  A similar mechanism has been previously suggested as the cause of the lunar Late Heavy Bombardment.  Thus, it may be that a large-scale dynamical instability, with more or less cataclysmic results, is an evolutionary step common to many planetary systems, including our own.

\end{abstract}
\section{Introduction}
\label{intro}
Currently, there are twenty-six detected multi-planet extrasolar systems (\url{http://exoplanet.eu/}).  Of these, at least eight  \citep{2007prpl.conf..685U} contain a pair of planets in a likely mean-motion resonance (MMR), wherein the planets' periods are maintained in an integer ratio 
(\citealt{1999..Murray..Dermott..book}) .  The stable locking of two bodies into such resonances requires a dissipative process that induces convergent migration between the pair.  Indeed, gravitational interaction with the gas nebula likely causes extensive migration in a young, forming planetary system 
(\citealt{1980ApJ...241..425G,1986ApJ...309..846L,1997Icar..126..261W}, see also \citealt{2007prpl.conf..655P} for a review); the establishment of MMRs is thus thought to be a consequence of this early evolution.  Several specific scenarios have been suggested.  The observed resonant exoplanets are all massive enough that they likely formed a fairly deep gap in their parent disk, thus being more or less locked into the disk's viscous evolution in what is referred to as type II migration.  Two gap-opening planets, if formed in close enough proximity, will clear out the intervening annulus of gas and so 
end up in a common gap 
\citep{2000ApJ...540.1091B,2000MNRAS.313L..47K}, or if the inner disk accretes faster than the planets migrate, at the inner edge of a disk cavity.  Subsequent capture into a mean-motion resonance is a very likely outcome 
\citep{2002ApJ...567..596L,2004A&A...414..735K}.  Differential migration will also tend to take place between gap-opening bodies co-evolving with the gas disk, and non-gap-opening bodies which usually migrate more rapidly (type I migration, 
\citealt{1997Icar..126..261W}); when the latter catch up to the former, capture into mean-motion resonances is again a likely result, as suggested by 
\cite{1996LPI....27..479H} and demonstrated by 
\cite{2005ApJ...626.1033T}.  A resonantly-captured smaller body may subsequently grow into a gas giant, providing another pathway to a pair of Jovian planets in a MMR.  In general, planet-disk interaction in a young planetary system may result in multiple planets, both gas giants and smaller, locked in MMRs.  

Such a picture leads naturally to the notion of planetary systems emerging from the gas disk era with crowded, compact architectures.  This, in turn, has been a recurring theme in formation models of planetary systems.  
\cite{1999Natur.402..635T,2002AJ....123.2862T} developed a model of giant planet formation in the Solar System wherein Uranus and Neptune originated in the same region ($\sim 5-10$ AU) as Jupiter and Saturn.  Proto-Jupiter's acquisition of a massive gas envelope then destabilized the closely-spaced system.  The ensuing scattering, combined with dynamical friction from the remaining outer planetesimal disk, then delivered the planets to their current orbits.  The model of 
\cite{2005Natur.435..466G} begins with a similarly compact configuration, but requires it to be stable until the time of the Late Heavy Bombardment (LHB), a cataclysmic event 700 Myrs after the initial formation of the Solar System, as implied by the Moon's cratering record 
(\citealt{1974E&PSL..22....1T,2000orem.book..493H} and references therein).  The instability which places the planets on their final orbits is then simultaneously invoked as the cause of the LHB.  Also, models for reproducing the  eccentricity distribution of the observed extrasolar planets by planet-planet scattering of course require the planets to start out close enough to each other for instability to ensue 
\citep{1996Sci...274..954R,1996Natur.384..619W,1997ApJ...477..781L,2007astro.ph..3166C,2007astro.ph..3160J}.  

Here, we combine these two notions and begin, in \S \ref{all_resonant_2 setup}, by constructing a model of a compact, resonantly-locked planetary system, as might plausibly be left behind by a dissipating gas disk.  Simulating the post-gas evolution, we then add an outer planetesimal disk beginning just beyond the outermost planet (\S \ref{all_resonant_2}).  Scattering of planetesimals induces planet migration, albeit on a much longer timescale than in the presence of a gas disk; 
this eventually drives the system to instability.  In \S \ref{all_resonant_3}, we demonstrate that this can lead to Solar System-like outcomes.  We explore two other scenarios, with more closely-spaced Jovian planets, in \S \ref{all_resonant} and \S \ref{all_resonant_Jouter}, and find that large gas giant eccentricities can be produced in the ensuing instability.  We discuss our results in \S \ref{discussion}.  

\section{A compact, resonantly-locked Solar System}
\label{all_resonant_2 setup}

\cite{2005ApJ...626.1033T} performed simulations of growing protoplanets exterior to a Jupiter-mass planet, and showed that a variety of resonant configurations can result.  The protoplanets end up occupying different exterior low-order resonances, and sometimes multiple planets end up sharing the same resonance; individual outcomes are stochastic.  Here, we wish to begin by constructing a version of the Solar System in which all the giant planets are (i) locked in MMRs with each other, and (ii) packed within, approximately, the current Jupiter-Saturn region.  

We want to do this in a way which mimics the action of planet-disk interactions.   We perform simulations with SyMBA 
\citep{1998AJ....116.2067D}, an N-body integrator optimized for near-Keplerian systems.  SyMBA uses the symplectic mapping of 
\cite{1991AJ....102.1528W}, with the addition of an adaptive timestep to resolve close encounters between pairs of bodies.  Bodies are merged when their separation is less than the sum of their physical radii.
We add to the code accelerations to model the radial migration and eccentricity damping due to gravitational interaction with the gas disk.  For type I migration, we use the result of 
\cite{2002ApJ...565.1257T}:
\begin{eqnarray}
t_{\rm migr} & \equiv &\frac{r}{|\dot{r}_{\rm migr}|}  \\
\, &=& \left (2.7 + 1.1 \beta \right )^{-1} \left ( \frac{M}{M_*}\right )^{-1} \left (\frac{\Sigma_g r^2}{M_*} \right )^{-1} \left ( \frac{H}{r} \right )^2 \Omega^{-1} \nonumber
\label{eq: Type I rate}
\end{eqnarray}
for a mass $M$ body orbiting a mass $M_*$ star at radius $r$, embedded in a gas disk with surface density $\Sigma_g$ and scale height $H$.  $\Omega$ is the Keplerian angular velocity and $\beta=-d \log \Sigma_g/d \log r$.  For eccentricity damping, we use
\begin{equation}
t_e \equiv \frac{e}{|\dot{e}|} =  \frac{1}{C_e} \left ( \frac {M}{M_*} \right )^{-1} \left (\frac{\Sigma_g r^2}{M_*} \right )^{-1} \left ( \frac{H}{r}\right )^4 \Omega^{-1}
\end{equation}
with $C_e \sim 1-10$ 
\citep{1989ApJ...345L..99W,1993Icar..106..274W,1993ApJ...419..166A,2000MNRAS.315..823P}.  
The corresponding acceleration applied within the code is $\vec{a} = a_{\phi} \hat{\phi} + a_r
\hat{r}$, where
\begin{equation}
a_{\phi} = -\frac{{v_{\phi}}}{2 t_{\rm migr}} 
\end{equation}
and
\begin{equation}
a_r = -2 \frac{\vec{v} \cdot \vec{r}} {r^2 t_e}.
\end{equation}
We then model type II (gap-opening) planets in a very simple way, as in 
\cite{2005ApJ...626.1033T}:  We choose a radius $r_{\rm
gap}$ for the center of the gap and a fractional gap ``width" $w$,
then for a type II planet with semimajor axis $a$, apply an azimuthal acceleration $a_{\phi}'$ as follows:
\begin{equation}
a_{\phi}'= \frac{a-r_{\rm gap}}{|a-r_{\rm gap}|+w a}a_{\phi}
\end{equation}
Since $a_{\phi}$ is negative, $a_{\phi}'$ is negative at $a>r_{\rm
gap}$, positive at $a<r_{\rm gap}$, and falls off to zero
as $a \rightarrow r_{\rm gap}$.  Therefore, if our ``gap-opening"
planet is displaced in semimajor axis from the gap midpoint, it experiences a
restoring torque towards $a=r_{\rm gap}$.  Although the eccentricity evolution of gap-opening planets is uncertain, for simplicity we calculate the eccentricity-damping part of the acceleration in the same manner:
\begin{equation}
a_{r}'= \frac{a-r_{\rm gap}}{|a-r_{\rm gap}|+w a}a_{r}
\end{equation}

We begin with a Jupiter-mass planet (310 M$_\oplus$) centered in its simulated gap at 5.5 AU.  A Saturn-mass body (95 M$_\oplus$), which is treated as being in the non-gap-opening regime, begins at 10 AU.
A Uranus and Neptune-mass planet are placed at 15 and 25 AU, respectively.  It should be noted that this configuration is not meant to necessarily reflect the initial state of the planetary system; rather, it is chosen as a simple way to allow one planet after another to migrate into resonance.  A more likely scenario---one which avoids a prohibitively long formation timescale---is that the planets originated much closer to Jupiter and Saturn \citep{1999Natur.402..635T,2002AJ....123.2862T} and
were locked into resonance as they grew 
\citep{2005ApJ...626.1033T}.  

\begin{figure}
\plotone{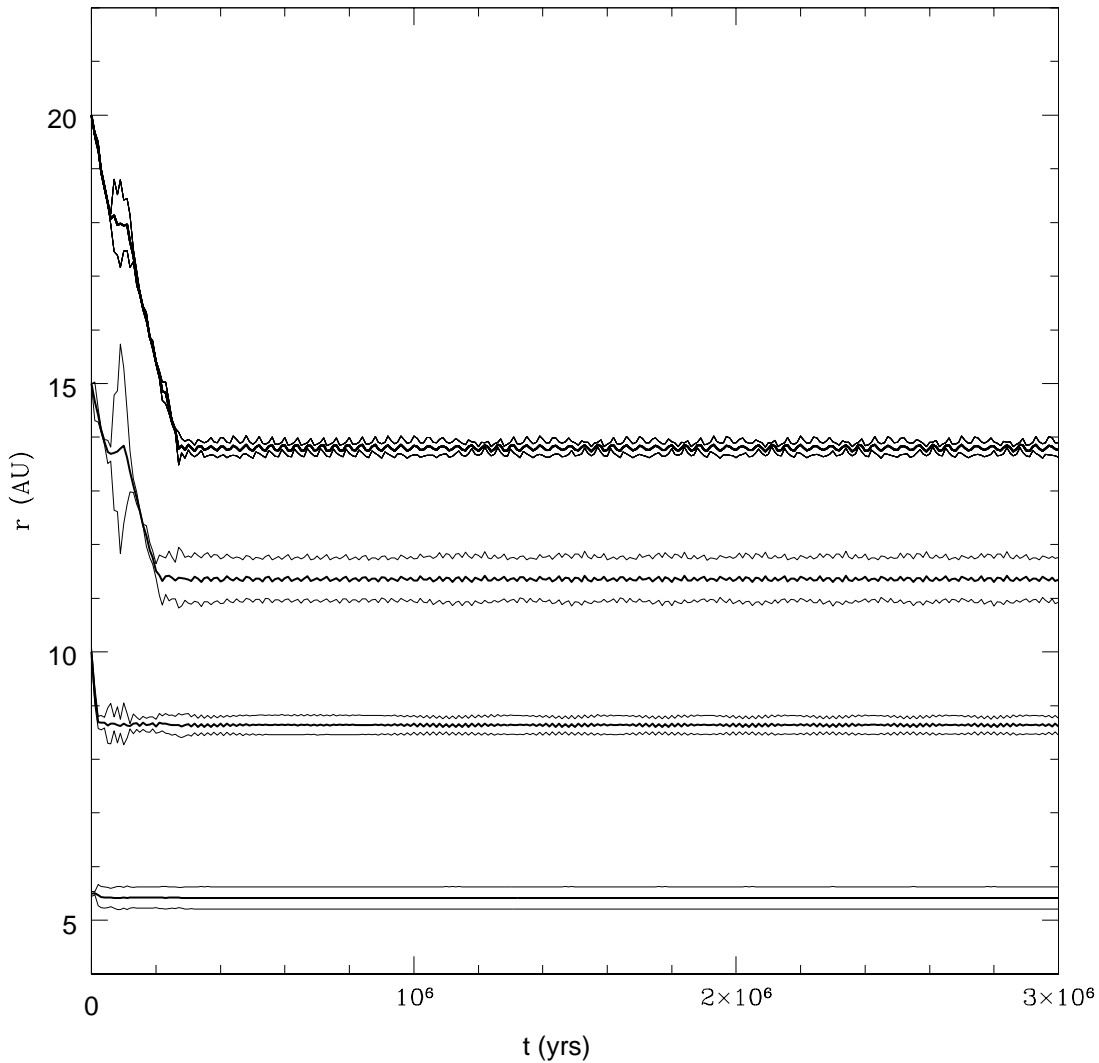}
\caption{Planetary migration due to simulated planet-disk interaction resulting in the establishment of a resonantly-locked version of the Solar System.  For each planet, pericenter distance, semimajor axis and apocenter distance are plotted as a function of time (all three being coincident for zero eccentricity). Jupiter and Saturn end up in a 2:1 mean-motion resonance (MMR), Saturn and Uranus in a 3:2 MMR, and Uranus and Neptune in a 4:3 MMR.}
\label{apa_PAPER_fig1}
\end{figure}

We adopt a gas disk scale height of 
\begin{equation}
H(r)=0.047 (r/{\rm 1 AU})^{5/4} {\rm AU}
\end{equation}
as in the model of 
\cite{1981PThPh..70...35H}.  We take the gas surface density to be of the form
\begin{equation}
\Sigma_g(r)=\Sigma_{g,{\rm AU}} \left ( \frac{r}{\rm 1 AU} \right )^{-1} e^{-t/10^6 {\rm yrs}}
\end{equation}
where the time-dependent exponential part models the observed dissipation of T Tauri gas disks on a Myr timescale 
\citep{2001ApJ...553L.153H}.  The outcomes show the same behavior described in 
\cite{2005ApJ...626.1033T}:  Higher surface mass densities produce stronger migration, and thus tend to lock bodies in closer MMRs.  At the same time, the value of the damping coefficient $C_e$ controls the ratio between migration and eccentricity damping, and thus affects the equilibrium eccentricity the bodies reach after being locked in resonance.  For $200 {\rm g cm^{-2}} \la \Sigma_{g,{\rm AU}} \la 1300 {\rm g cm^{-2}}$ and $C_e=1$, we obtain a stable configuration in which Jupiter and Saturn are in a 2:1 MMR, Saturn and Uranus are in 3:2, and Uranus and Neptune are in 4:3, all with eccentricity $\la 0.05$.  This is shown in Fig. \ref{apa_PAPER_fig1}.  It should be pointed out that these low eccentricities result because eccentricity damping is applied to {\it all} the planets, under the assumption that all of them are within the  disk.  In contrast, when damping is only applied to one of a pair of resonant planets---as would happen when the inner migrates within a disk cavity---the eccentricities of both will become large unless damping on the outer is very strong \citep{2002ApJ...567..596L,2003ApJ...597..566T}.

\section{Breaking apart a resonantly-locked Solar System}
\label{all_resonant_2}

We numerically evolve different versions of this system (differing in the initial orbital angles of the planets) to $10^9$ years, with the planet-disk interaction decreasing on a timescale of $10^6$ years to simulate the dispersal of the gas disk; all remain completely stable.  Resonant capture due to disk-induced migration thus provides a natural way of producing planetary systems which are not only close-packed but also long-lived.  In contrast, comparably compact but non-resonant versions of this system require significant fine-tuning to assemble if they are not to go unstable on timescales $\ll 10^6$ years.  

\begin{figure}
\plotone{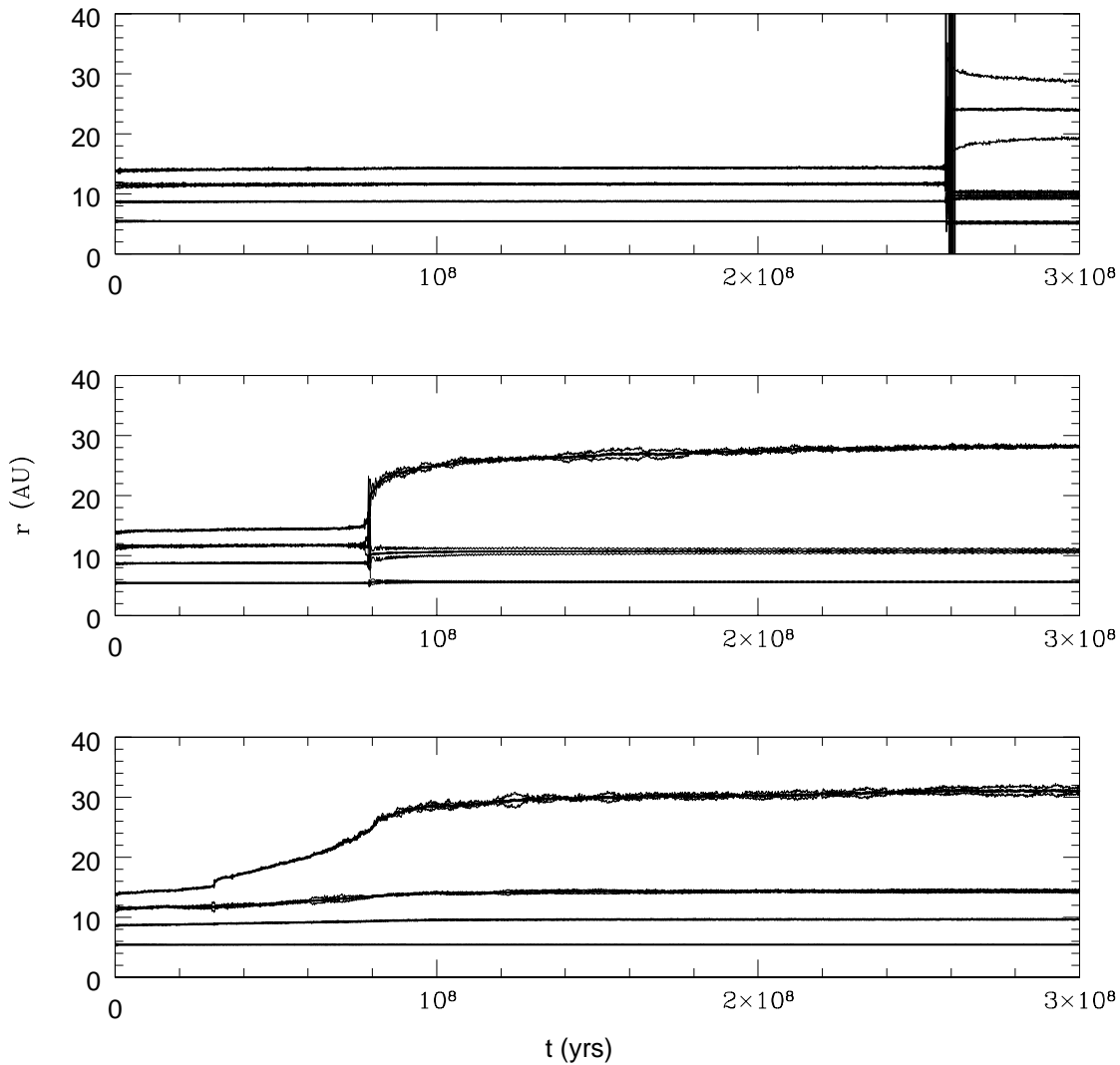}
\caption{Three simulations out of a set of 30, showing the evolution of an initially compact, resonant-locked version of the Solar System, as described in \S \ref{all_resonant_2 setup}, interacting with an outer planetesimal disk.  Semimajor axis as well as peri- and apocenter distance are plotted as a function of time.  In the first two panels, one of of the Uranus/Neptune mass planets is abruptly ejected after $2.6 \times 10^8$ years and $8 \times 10^7$, respectively.  In the last panel, although the ice giants clearly receive a kick at $3 \times 10^7$ years, strong scattering does not take place, and the planets primarily just evolve by scattering planetesimals; ``Neptune" does not stop until it reaches the outer planetesimal disk edge at 30 AU.}
\label{run2_run11_run19_PAPER_fig2}
\end{figure}

By themselves, then, these resonant systems are likely to remain essentially ``frozen" in the configuration with which they emerge from the gas.  However, in reality our systems would have an important additional component, namely the remnant outer planetesimal disk, where significant planet growth did not have time to occur before the disappearance of the gas \citep{2003Icar..161..431T}.
Scattering of leftover planetesimal very likely drove divergent migration of the giant planets in the early Solar System as follows 
\citep{1984Icar...58..109F,1999AJ....117.3041H,2004Icar..170..492G}:  Planetesimals from the inner disk edge are perturbed onto orbits that cross the planets.  Planetesimals are then scattered from planet to planet until they are ejected.  When the innermost planet is much more massive than the others (as is Jupiter), it is most efficient at ejecting planetesimals.  In the process, the massive planet loses angular momentum and migrates inward by a small amount, while the smaller outer planets, as they pass the planetesimals inward, undergo a net gain in angular momentum and migrate outwards.  For given planetary masses, the rate of migration increases with the surface density of planetesimals; at a high enough density, ``runaway" migration may even result 
\citep{2004Icar..170..492G}.  The numerous Kuiper belt objects in exterior MMRs with Neptune---including Pluto and its fellow ``Plutinos" in the 3:2 resonance---are thought to be the result of resonant capture during this migration.  
It is reasonable to expect that this same mechanism would act to pull apart our resonantly locked planets.  How would this affect the dynamics of the system?
To investigate, we perform a series of simulations, taking the resonant system assembled in the previous section and adding an outer planetesimal disk.  The  disk initially extends from 15.3 AU (1.5 AU beyond the outermost planet) to 30 AU.  The rather small extent of the disk is motivated 
by the models of 
\cite{2003Natur.426..419L} and \cite{2005Natur.435..466G}. 
These invoke an initially compact planetesimal disk, truncated at $\sim 30 $AU, with the present-day Kuiper belt consisting of bodies later transported outward, and Neptune's migration through the planetesimal disk having been stopped simply by reaching the outer disk edge.  The individual planetesimals have a mass of 0.035 M$_\oplus$, much larger than physically realistic planetesimals, but small enough compared to the giant planets that the key effects we are interested in---planetesimal-driven migration and dynamical friction---are adequately modeled.  Planetesimals are treated as ``second-class" particles, which interact fully with the four planets but not with each other.    
The planetesimals are distributed with a surface density 
\begin{equation}
\Sigma_{\rm plsml}=\Sigma_{\rm plsml,AU} \left ( \frac{r}{\rm 1 AU}\right )^{-1}.
\label{plsml disk}
\end{equation}
We perform a set of 30 simulations, linearly varying $\Sigma_{\rm plsml,AU}$ from 4 to 16 ${\rm g\,cm^{-2}}$ (corresponding to a total planetesimal disk mass of about 14 to 55 M$_\oplus$, made of of 400 to 1600 planetesimals).  Each runs to $3 \times 10^8$ years.  In all but two of the simulations, the orbits of the giant planets undergo significant dynamical changes.  Fig. \ref{run2_run11_run19_PAPER_fig2} shows three representative cases.  
The evolution is particularly dramatic in the first two of these, both of which undergo a scattering event that results in the ejection of one of the Uranus/Neptune-mass bodies.  This behavior---a long period of slow, quiescent evolution followed by abrupt instability---is reminiscent of what occurs in the scenario of 
\cite{2005Natur.435..466G}.  There, an instability is triggered when Jupiter and Saturn, migrating divergently, cross their 2:1 MMR.  Divergent resonance passage cannot result in resonant capture, but does give a ``kick" in eccentricity to the two bodies involved, with closer and lower-order resonances having a stronger effect 
\citep{1988Icar...76..295D,2002ApJ...564L.105C}.
\cite{2005Natur.435..466G} show that the abrupt increase in Jupiter and Saturn's eccentricity in a compact version of the Solar System can cause Uranus and Neptune to be strongly scattered.  Similarly to the model of 
\cite{1999Natur.402..635T}, this can propel them toward their wider current orbits, while dynamical friction with the planetesimal disk serves to damp their eccentricities.  This large-scale shakeup of the Solar System also provides a plausible source for the Late Heavy Bombardment, by causing the terrestrial region to suffer a sudden increase in the flux of small bodies, both cometary (from the outer disk) and asteroidal (putative bodies perturbed from thus far stable orbits in the Asteroid Belt).  

\begin{figure}
\plotone{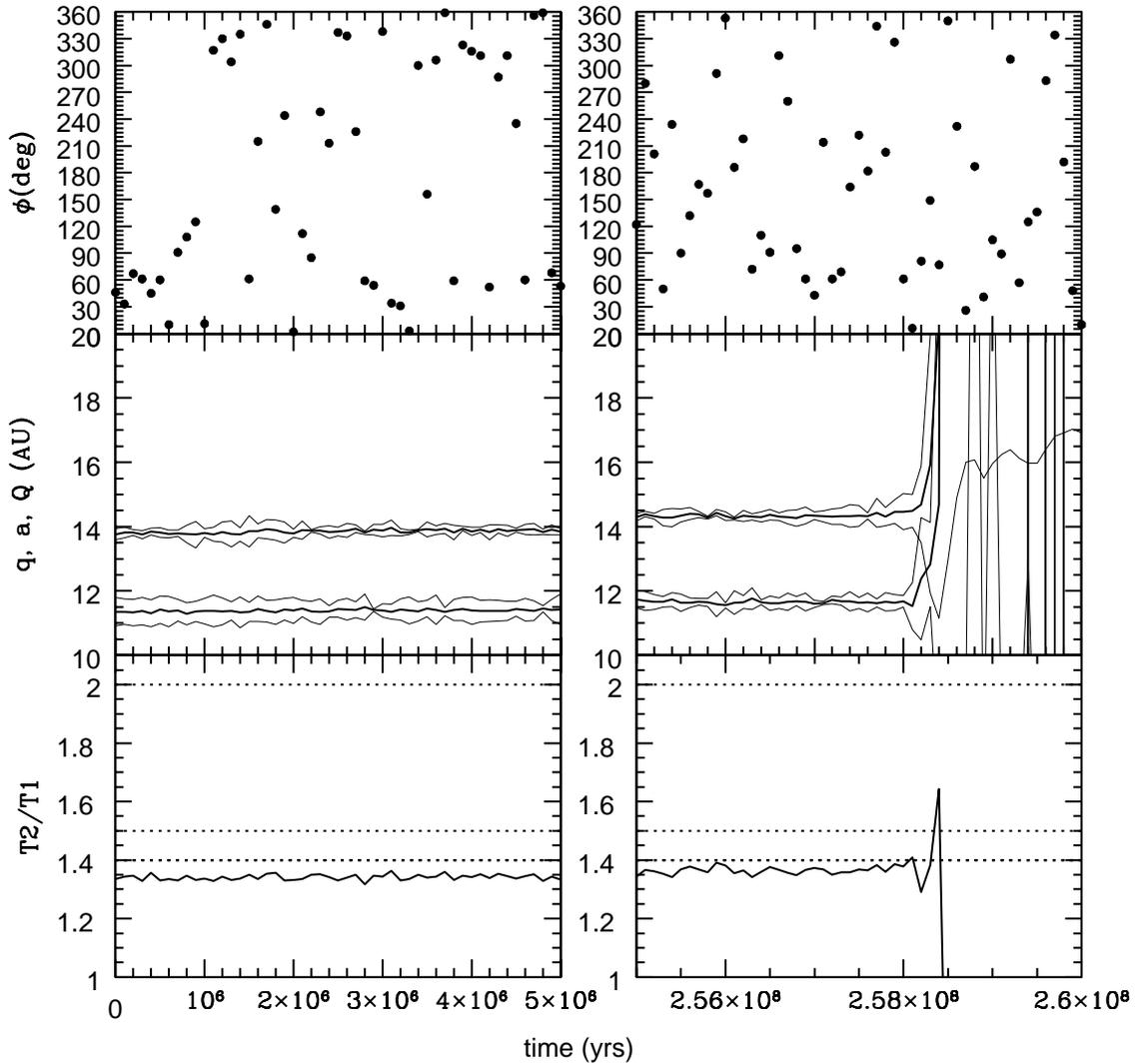}
\caption{A closer look at the evolution of the outer two ice giants in one of the simulations of \S \ref{all_resonant_2}, showing $\phi_?$, one of the 4:3 resonance angles (top panel), the pericenter ($q$), semimajor axis ($a$) and apocenter ($Q$) distances of the two planets (middle panel), and their period ratio (bottom panel) for the first 5 Myrs, and around the time the system becomes unstable.  This happens as the two planets cross their 7:5 mean-motion resonance ($T_2/T_1=1.4$).} 
\label{run2_closer_look_PAPER_fig3}
\end{figure}

However, this process cannot be the one responsible for the instabilities we find here, for a simple reason:  Jupiter and Saturn are {\it already} in a 2:1 MMR.  In fact, we find that they do not leave this resonance until the moment the instability sets in.  Thus, the trigger for the instability must be something else.  Fig. \ref{run2_closer_look_PAPER_fig3} gives a closer look at what happens.  ``Uranus" and ``Neptune" are initially in a 4:3 MMR, as assembled in \S \ref{all_resonant_2 setup}.  Interaction with the planetesimal disk pulls them apart, and they soon leave the resonance, signified by the 4:3 resonance angles,
\begin{equation}
\phi_1= 4 \lambda_2 -3 \lambda_1 - \tilde{\omega_1} {\rm \,\,\,and\,\,\,} \phi_2= 4 \lambda_2 -3 \lambda_1 - \tilde{\omega_2}
\end{equation}
(see e.g. \citealt{1999..Murray..Dermott..book}) switching from libration to circulation, where $\lambda$ are the planets' instantaneous longitudes, $\tilde{\omega}$ their longitudes of pericenter, and $1,2$ denote the inner and outer planet, respectively.  This happens after only $\sim 2$ Myrs, but the system continues to be stable.  It is much later that the instability sets in, and it happens when Uranus and Neptune divergently cross their 7:5 MMR and receive an impulsive increase in their eccentricities.  This increase is not by itself large; separate simulations with only Uranus and Neptune and artificially-imposed migration on a similar timescale to what is induced by planetesimal scattering show that $\Delta e$ is always far below 0.1 for both planets.  If the system consists only of these two planets, no instability results, as can be seen by considering the Hill stability criterion for the minimum separation of a pair of equal-mass ($M/M_*=\mu$) eccentric planets:
\begin{equation}
\frac{\Delta a}{a} \ga \sqrt{\frac{8}{3}(e_1^2 + e_2^2)+9 \mu^{2/3}}
\end{equation}
where $\Delta a/a$ is the fractional orbital separation \citep{1993Icar..106..247G}.  For Uranus/Neptune mass bodies orbiting a Solar-mass star, $\mu \approx 5 \times 10^{-5}$, and for the 7:5 resonance, $\Delta a/a=(7/5)^{2/3}-1=0.25$; assuming $e_1 \approx e_2$, the pair remains stable unless $e_1$ and $e_2$ are excited to 0.1 by the resonance passage.   

However, with Saturn and Jupiter nearby, this modest eccentricity increase turns out to be sufficient to initiate a chain reaction of instability, with both ice giants crossing the gas giants' orbits.  One of the ice giants is ejected from the system.  
At the same time, Jupiter and Saturn receive a sufficiently strong kick that they are knocked out of their 2:1 MMR, scattering each other and ending up about 1 AU further apart with higher eccentricities, $\approx 0.04$ and 0.06, respectively.  

A very similar evolution takes place for about a third of the simulations.  The trigger for instability is the divergent passage of Uranus and Neptune through either their 7:5 or 3:2 MMR, and the kick in eccentricity this interaction administers to both planets.  Though the exact outcomes vary stochastically from simulation to simulation, there is a clear overall correlation with the planetesimal disk surface density, with higher disk densities decreasing the strength of the scattering.  This is because higher-mass disks produce faster planetesimal-driven migration, resulting in faster passage through MMRs and thus a weaker impulse for Uranus and Neptune.  At the same time, dynamical friction with the disk is stronger, so that whatever eccentricity the planets do manage to pick up is damped more effectively.  As a result, none of the cases with $\Sigma_{\rm plsml,AU} \ga 8 {\rm g\,cm^{-2}}$ result in scattering.  Instead, all of them display variants of the behavior seen in the bottom panel of Fig. \ref{run2_run11_run19_PAPER_fig2}:  The outermost planet migrates outward through the planetesimal disk at low eccentricity, not stopping until it arrives near the original outer edge of the disk (30 AU).  These bodies are in fact undergoing runaway or ``forced" migration, as detailed in 
\cite{2004Icar..170..492G}. 

\begin{figure}
\plotone{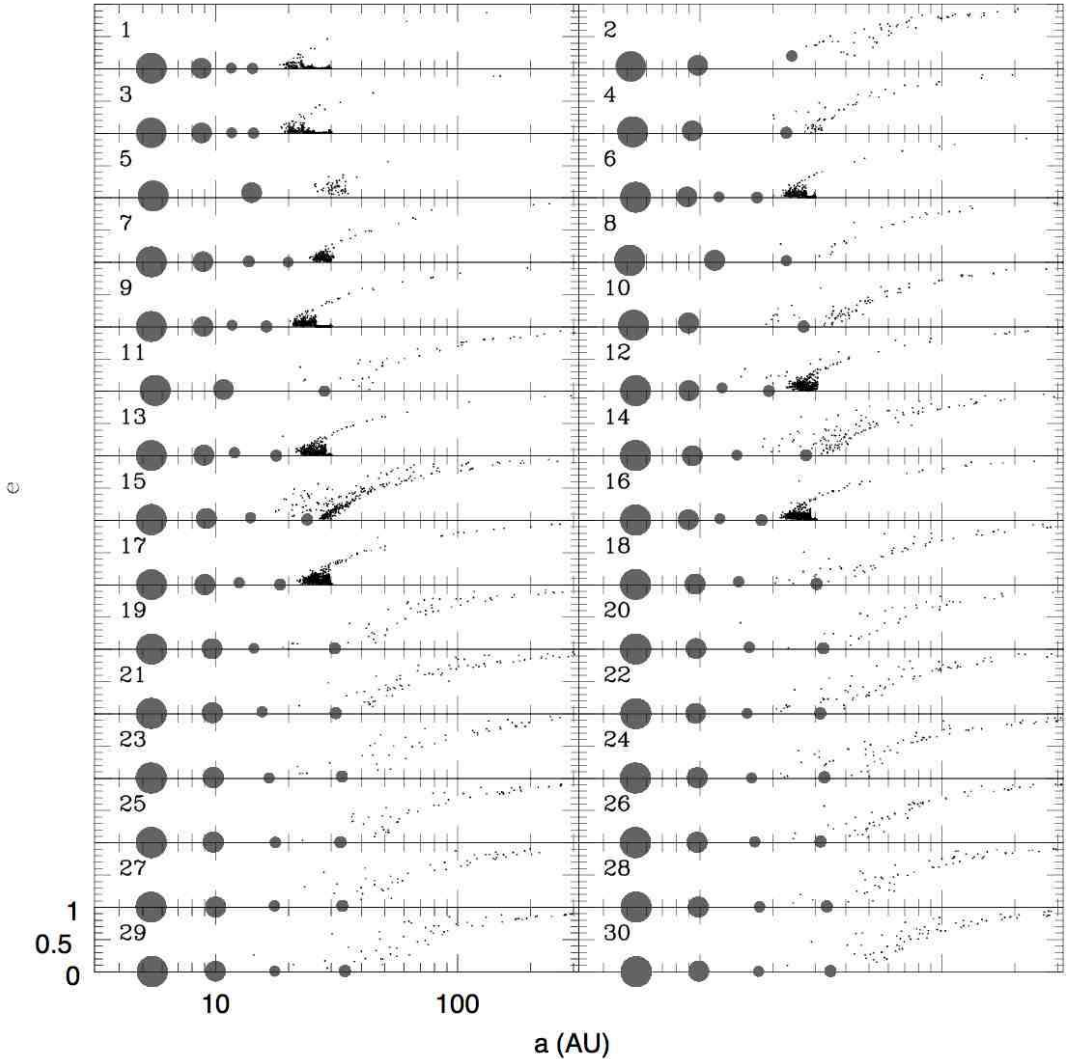}
\caption{The state of the 30 simulations in \S \ref{all_resonant_2} at their stopping time of 300 Myrs.  Eccentricity vs. semimajor axis (averaged over the last $10^6$ yrs) is plotted for the planets (gray, size $\propto$ physical size) and the planetesimals.  Initial planetesimal disk surface densities are given by Eq. \ref{plsml disk}, with $\Sigma_{\rm plsml,AU}$ linearly increasing from 4 to 16 ${\rm g\,cm^{-2}}$ from Run 1 to Run 30.
Broadly speaking, the systems evolve in one of three ways: (i) Almost no change in the giant planet orbits over 300 Myrs (Runs 1 and 3); (ii) divergent migration of the outer planets, resulting in sudden instability and scattering when they cross their 7:5 or 3:2 MMR (Runs 2,4,5,7,8,10 and 11); (iii) divergent migration in which resonance crossing does not destabilize the system, in the sense that no orbits cross (all remaining runs). }
\label{frames_onepage_linlog_PAPER_fig4}
\end{figure}

Snapshots of all 30 simulations at their completion time of $3 \times 10^8$ years are shown in Fig. \ref{frames_onepage_linlog_PAPER_fig4}.  Insofar as we are looking for an analog to the Solar System, the outcomes reveal somewhat of a ``catch-22":  All of the cases in which strong scattering occurs lose one of the ice giants.  On the other hand, in the cases with higher disk masses, in which planetesimal-driven migration dominates, the absence of an abrupt instability does away with the desirable Late Heavy Bombardment part of the picture.  Also, when the ice giants' orbital evolution is predominantly the result of migration, the mechanism works much less efficiently for the inner of the two ice giants.  Therefore ``Uranus" is systematically placed at too small an orbital radius in the end, a dilemma pointed out by 
\cite{2004Icar..170..492G}.  This effect is clearly visible in Fig. \ref{frames_onepage_linlog_PAPER_fig4}. 

\section{Solar System-like outcomes}
\label{all_resonant_3}
\begin{figure}
\plotone{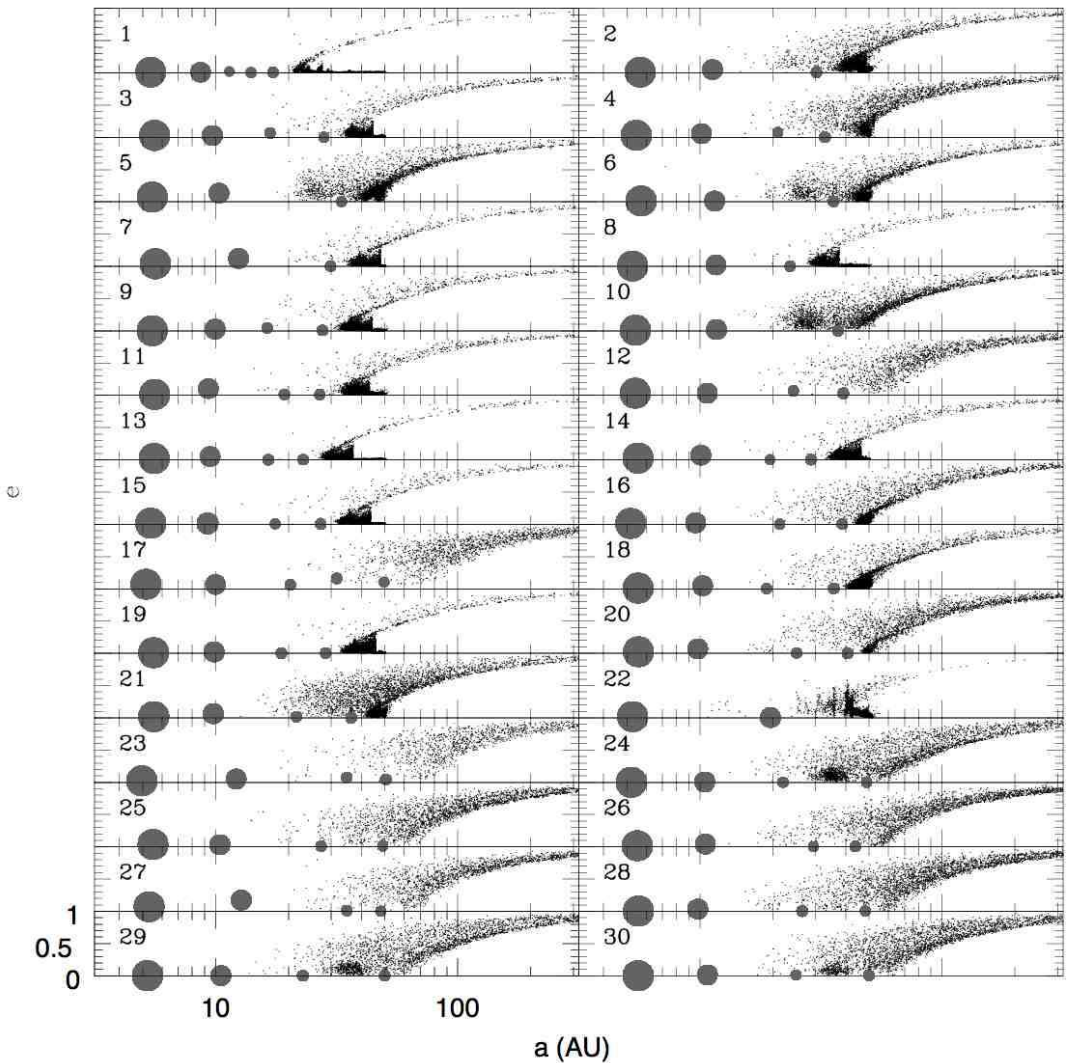}
\caption{The state of the 30 simulations in \S \ref{all_resonant_3} at their stopping time of 200 Myrs.  Eccentricity vs. semimajor axis (averaged over the last $10^6$ yrs) is plotted for the planets (gray, size $\propto$ physical size) and the planetesimals.  Initial planetesimal disk surface densities are given by Eq. \ref{plsml disk}, with $\Sigma_{\rm plsml,AU}$ linearly increasing from 4 to 8 ${\rm g\,cm^{-2}}$ between Run 1 and Run 30. }
\label{frames_onepage_linlog_PAPER_fig5}
\end{figure}

We have demonstrated that interaction with an outer planetesimal disk will tend to eventually pull apart a compact, resonant version of the Solar System.  For lower-mass planetesimal disks, the most important part of this evolution is usually an abrupt instability triggered when ``Uranus" and ``Neptune", having left their original MMR, encounter a more distant one.  For higher disk masses, this instability tends to be suppressed or absent, and the giant planets' orbits evolve by planetesimal-driven migration alone.  We will focus on the former scenario.  To this end, we conduct another set of simulations and begin by choosing a lower range of disk masses.  Again distributing the planetesimals as per Eq. \ref{plsml disk}, we now let $\Sigma_{\rm plsml,AU}$ range from 4 to 8 ${\rm g\,cm^{-2}}$.  An added consequence of adopting such a relatively low disk mass is that planetesimal-driven migration is more likely to be in the ``damped" regime 
\citep{2004Icar..170..492G}, wherein migration stops by iteself after a few AU.  Thus an outer disk edge at $\sim 30$ AU should no longer be critical to stopping Neptune at the right place.  To test this, we extend the outer disk edge to 50 AU.  The disk masses thus range from about 31 to 63 M$_\oplus$.  Having noted that all instances of strong scattering in \S \ref{all_resonant_2} resulted in the loss of one of either ``Uranus" or ``Neptune", we simply add an extra ice giant in Neptune's exterior 4:3 MMR.  
With initial exploratory runs giving encouraging results, we increase the fidelity of the simulations by decreasing the planetesimal masses to $5 \times 10^{-3}$ M$_\oplus$ (less than half a Lunar mass), so that the lowest-mass planetesimal disk is modeled with $\approx 6200$ particles, and the highest-mass one with $\approx 12500$.  Due to the significantly increased computational cost, we only run to $2 \times 10^8$ years.  Snapshots of the simulations at this time are shown in Fig. \ref{frames_onepage_linlog_PAPER_fig5}.

All except one of the simulations (run 1) undergoes a scattering instability within this time.  In most cases, this is the result of the inner and middle ice giants crossing the 7:5 or 3:2 MMR as they diverge, though in some cases it is the middle and outer.  The important point is that even with this different initial configuration, it continues to be the small outer planets which served as the trigger for the instability.  

Eight of the outcomes resemble the Solar System in the sense that two ice giants are left with low eccentricities inside $\sim$ 40 AU, and are undergoing little or no migration: runs 3, 9, 11, 12, 13, 14, 15, 19 fall into this category.  Another four cases, runs 4, 16, 18 and 21, look like reasonable Solar System analogs but are still undergoing substantial migration, and are thus likely to end up with their ``Uranus" and ``Neptune" significantly more widely spaced than in the Solar System if their evolution is followed beyond $2 \times 10^8$ years.  Finally, runs 23 and higher have a sufficiently massive disk that planetesimal migration subsequent to the instability has driven the outermost planet to end up at or near the original outer disk edge at 50 AU.  It is also worth noting that of the Solar System-like outcomes above, all except 12 end up with at least a part of their outer disk largely pristine.  Thus, an additional depletion mechanism, such as collisional grinding 
(e.g. \citealt{1997ApJ...490..879S,1997Icar..125...50D}) would need to act in order to reproduce the inferred low mass of the present-day Kuiper belt, $\la 0.1$ M$_\oplus$.  

\section{Other initial conditions }
\subsection{A more compact system} 
\label{all_resonant}
\begin{figure}
\plotone{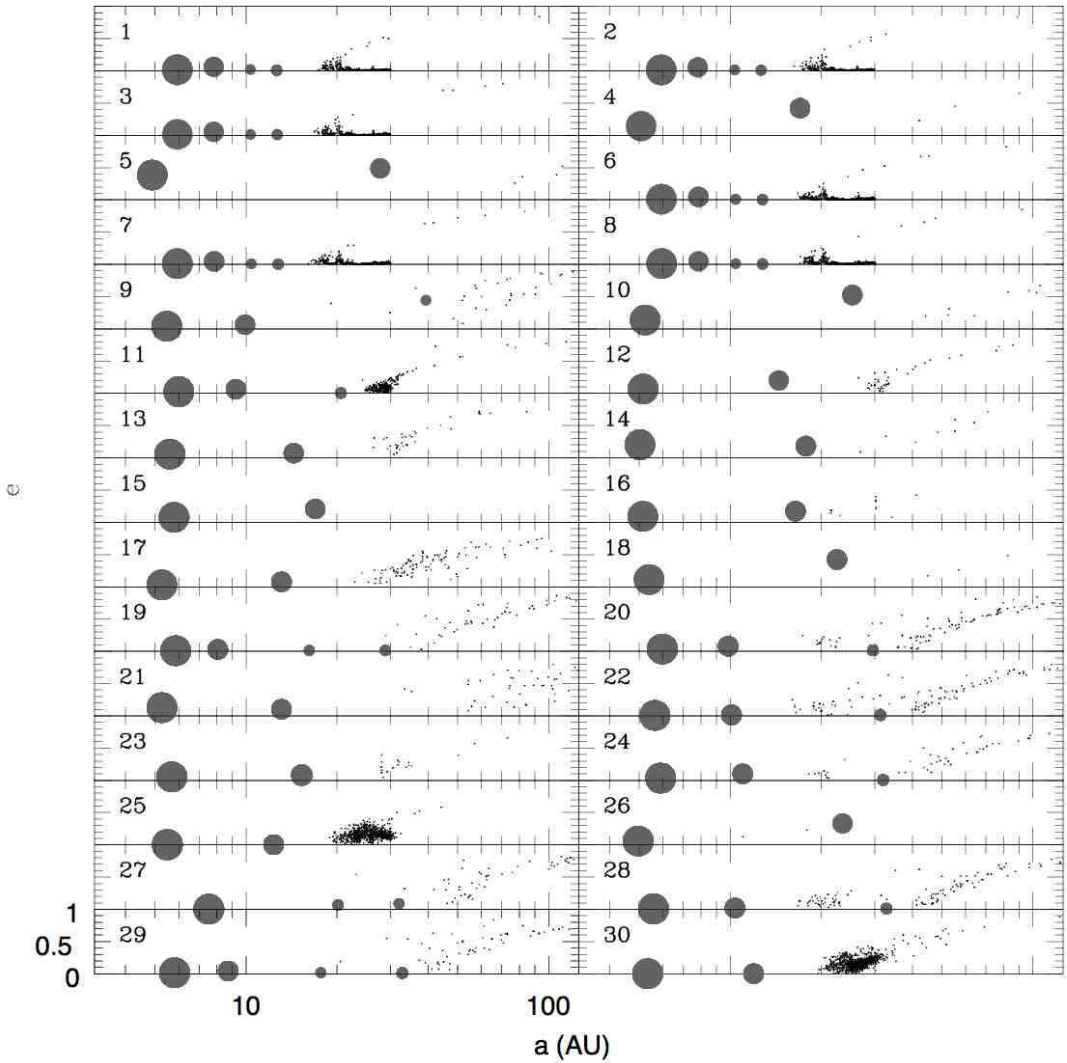}
\caption{The state of the 30 simulations in \S \ref{all_resonant} at their stopping time of 300 Myrs.  Initial conditions are as in \S \ref{all_resonant_2} but with Jupiter and Saturn beginning in a 3:2 instead of a 2:1 MMR.  Eccentricity vs. semimajor axis (averaged over $10^6$ yrs) is plotted for the planets (gray, size $\propto$ physical size) and the planetesimals.  Initial planetesimal disk surface densities are given by Eq. \ref{plsml disk}, with $\Sigma_{\rm plsml,AU}$ linearly increasing from 4 to 8 ${\rm g\,cm^{-2}}$ between Run 1 and Run 30. }
\label{frames_onepage_linlog_PAPER_all_resonant}
\end{figure}

We now explore the evolution of a planetary system from a few other initial configurations, though for simplicity we keep the Solar System giant planets as our ``building blocks".
We begin with a resonant system assembled in the same way as in \S \ref{all_resonant_2 setup}, except that Jupiter and Saturn start between a 2:1 and 3:2 period ratio.  As a result, the two gas giants are captured into the 3:2 MMR.  
We again add an outer planetesimal disk.  We revert to the coarser, smaller disk of \S \ref{all_resonant_2}, made up of 0.035 M$_\oplus$ planetesimals and with an outer edge at 30 AU.  Another set of 30 simulations (to $3 \times 10^8$ years)  is performed, with the same range of disk surface densities as in \S \ref{all_resonant_2}.  Fig. \ref{frames_onepage_linlog_PAPER_all_resonant} shows that we now obtain a number of systems in which one or both gas giants have eccentric orbits.  This is in sharp contrast to the simulations performed in \S \ref{all_resonant_2} and \S \ref{all_resonant_3}:  there, although the instability typically administers a strong enough kick to throw the gas giants out of resonance, they never acquire large eccentricities.  
However, with the two largest planets now starting out in the closer 3:2 MMR, the situation changes, and they strongly scatter each other in several cases.  In almost a quarter of the runs---4,5,10,12,14,18 and 26---one or both gas giants end up with substantial eccentricities ($\ga 0.2$).   In one case, Run 27, ``Saturn" is lost from the simulation.  However, this is not the result of an ejection; it happens when the planet crosses the inner simulation domain boundary at 2 AU, and so constitutes a rather artificial result.  Also worth noting is that ``Saturn" is in some cases scattered to a significantly larger semimajor axis, as high as $\approx 30$ AU.  
\newpage
\subsection{Reversing Jupiter and Saturn}
\label{all_resonant_Jouter}

\begin{figure}
\plotone{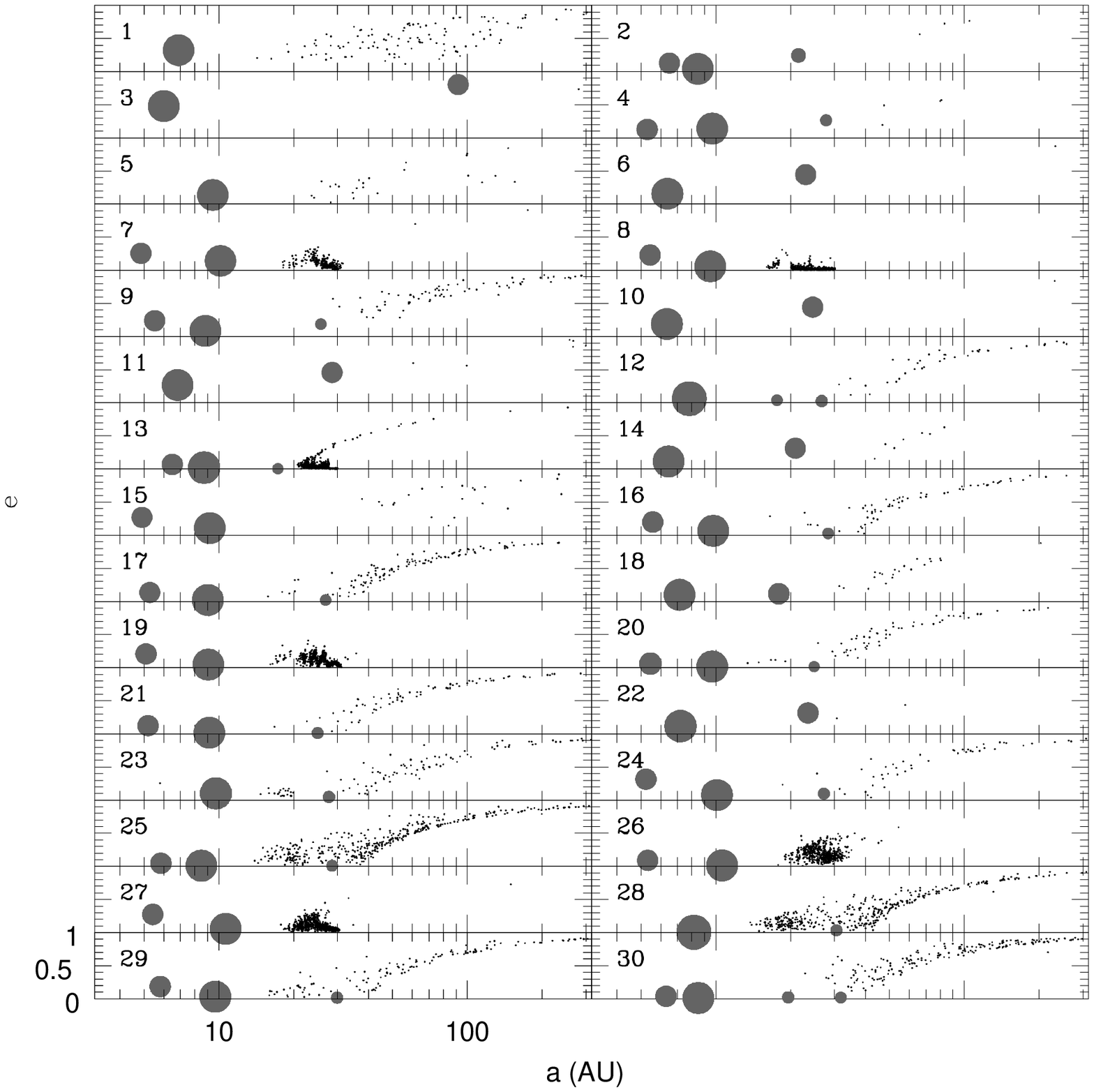}
\caption{The state of the 30 simulations in \S \ref{all_resonant_Jouter} at their stopping time of 300 Myrs.  Initial conditions are as in \S \ref{all_resonant} but with Jupiter and Saturn reversed, so that Saturn is the innermost planet. Eccentricity vs. semimajor axis (averaged over $10^6$ yrs) is plotted for the planets (gray, size $\propto$ physical size) and the planetesimals.  Initial planetesimal disk surface densities are given by Eq. \ref{plsml disk}, with $\Sigma_{\rm plsml,AU}$ linearly increasing from 4 to 8 ${\rm g\,cm^{-2}}$ }
\label{frames_onepage_linlog_Jouter_PAPER}
\end{figure}

The model of 
\cite{2005Natur.435..466G} requires Jupiter and Saturn to migrate apart and cross their 2:1 MMR, something which is only possible because the inner gas giant, being more massive, is more efficient at ejecting planetesimals.  In contrast, the instability mechanism we have examined here only requires the divergent migration of the small outer Uranus/Neptune-mass planets, and thus ought to function independently of the gas giant mass ratio.  As a demonstration, we assemble the resonant configuration of \S \ref{all_resonant} above, except that we reverse the ordering of Jupiter and Saturn, making the latter the innermost planet.  The results are shown in Fig. \ref{frames_onepage_linlog_Jouter_PAPER}.  In contrast to the previous sets of simulations, here {\it all} have undergone strong scattering within 300 Myrs.  
The higher yield of instability is the result of the smaller outer planets now having a (three times) more massive neighbor.  This lowers the threshold for how strong a perturbation to the outer planets' orbits is needed in order to trigger global instability.  In comparison to the previous set of simulations in \S \ref{all_resonant}, we also produce a larger fraction of cases---one half---in which at least one of the gas giants ends up with an averaged eccentricity $\ga 0.2$ (runs 1, 3, 6, 7, 8, 9, 10, 11, 14, 15, 16, 19, 22, 24, 27). In one of these cases (1) ``Saturn" is ejected from the system, and in another (3) it acquires a semimajor axis of almost 100 AU.  Among the remaining lower-eccentricity outcomes, there are two instances (12, 28) of a physical collision, and thus a merger, between the two gas giants.  In another two cases (5, 23) ``Saturn" is removed when it crosses the inner boundary.   
\newpage
\section{Discussion and Conclusions}
\label{discussion}
We have shown that migration in a young protoplanetary disk can readily produce systems of planets in which each member is locked in a mean-motion resonance (MMR) with its neighbors.  Due to the stabilizing effect of the resonances, even tightly-packed configurations, with period ratios of adjacent planets ranging from 2:1 to 4:3, are stable over timescales long compared to the gas disk lifetime ($10^6$ to $10^7$ years), even after the dissipational effect of the gas is removed.  We have then gone on to show that at later times such configurations can be destabilized, frequently in a catastrophic manner involving strong planet-planet scattering.  This requires divergent planet migration, which can by driven be the interaction with an outer planetesimal disk.  The actual trigger is a pair of planets crossing a mutual MMR, which for diverging orbital periods produces eccentricity excitation but not capture.
A key feature we find is that in a compact system of Jupiter/Saturn-mass inner planets combined with much smaller Uranus/Neptune-mass outer planets, the latter alone can serve as the trigger for global instability, a case of the ``tail wagging the dog".  

Reverse resonance crossing was invoked as the trigger for the scattering of Uranus and Neptune, and simultaneously for the Late Heavy Bombardment, in the model of 
\cite{2005Natur.435..466G}, but in contrast, they require the largest planets, Jupiter and Saturn, to cross a MMR (the 2:1).  Also, no planets are initially in resonance \footnote{As we were writing this paper, it was brought to our attention that a new version of their model also begins with the giant planets in MMRs (Morbidelli et al., in preparation)}.
The problem is that planetesimal-induced migration moves the less-massive Uranus and Neptune much faster than it does Jupiter and Saturn, yet at the time of resonance crossing, the system still needs to be compact enough that the ice giants have a high probability of being scattered.  In order that they cross the resonance quickly enough, Jupiter and Saturn must therefore start out just a bit closer than the 2:1.  Thus, the system must have emerged from the gas disk in a rather finely-tuned configuration, made even more precarious by the lack of any stabilizing MMRs between the closely-packed planets.  However, notwithstanding this issue,
\cite{2005Natur.435..466G} demonstrate that the onset of planetesimal-driven migration can be delayed by at least $10^9$ years (and in principal arbitrarily long) depending on how far the inner edge of the planetesimal disk is from the outermost planet.  It is this feature---the ability to initiate an instability after a long delay---which makes this mechanism a good candidate for triggering the Late Heavy Bombardment.  In the simulations presented here, instability generally sets in on a timescale $\la 10^8$ years; since we perform many simulations, we avoid a prohibitive computational cost by placing the inner disk edge close enough to produce a relatively rapid onset of migration.  

Our goal here has not been to undertake a thorough parameter study of outcomes produced by the breakup of resonant planetary systems; rather, we have tried to present a few interesting cases---simply from mixing and matching the Solar System giant planets---to illustrate the key features of this mechanism, and to serve as a jumping-off point for future work.  We have identified one pathway for producing systems which resemble the Solar System, beginning with a moderately compact set of planets: Jupiter and Saturn in a 2:1 MMR, and three exterior resonant Uranus/Neptune-mass planets, at least one of which is usually lost when the system becomes unstable.  We find that if Jupiter and Saturn start in a closer 3:2 MMR, strong scattering between the two gas giants can occur when instability sets in. In almost one quarter of this set of simulations, either Saturn, or both Jupiter and Saturn, are left with substantial eccentricities as a result.   Instability continues to take place---in fact the fraction of cases left with an eccentric gas giant rises to one half---when the orbits of Jupiter and Saturn are switched.  Therefore, this 
mechanism also provides a pathway to producing the sort of high eccentricities possessed by a large fraction of observed exoplanets.  Planet-planet scattering was proposed as a way of generating eccentricities as far back as the discovery of the first exoplanets 
\citep{1996Sci...274..954R,1996Natur.384..619W,1997ApJ...477..781L}, and recent results on reproducing the observed exoplanet eccentricity distribution in this way look promising
\citep{2007astro.ph..3166C,2007astro.ph..3160J}.  The scenario we develop here amounts to a means of systematically producing scattering from an initially stable configuration.  Again, though, it should be emphasized that we have only considered a few specific cases; the inclusion of a planetesimal disk renders the problem more dynamically complex (and much more computationally expensive) than a few-body scattering problem, thus significantly more work is required in order to extract an eccentricity distribution from this model which can be directly compared to observations.  
Likewise, we cannot say much about the semimajor axis distribution, except to note that in starting with the gas giants between $\sim 5-10$ AU, we do not produce any close-in giant planets (although our simulations have an inner boundary at 2 AU, there are only very few instances of a gas giant crossing it).  This is consistent with earlier findings that planet-planet scattering by itself is probably not able to account for hot Jupiters \citep{2001Icar..150..303F}.  On the other hand, in a number of cases the smaller of our gas giants (``Saturn") is scattered far outward, in one instance acquiring a semimajor axis of nearly 100 AU.  

With the above qualifications, our results suggest the possibility that the violent breakup of close-packed, resonantly-locked planets is an evolutionary step that has occurred in many planetary systems.  The exoplanets observed to be in MMRs would then represent simply the survivors of a much larger primordial 
resonant population.  
The Late Heavy Bombardment in our own Solar System may have actually resulted from a relatively gentle version of such a breakup, with more violent outcomes recorded in the high eccentricities common among observed exoplanets.  Indeed, recent Spitzer observations suggest that extrasolar versions of the LHB could be commonplace.  
\cite{2007ApJ...658..569W} find that 2\% of Sun-like stars exhibit hot dust in what corresponds to the terrestrial planet region; for most of these, the luminosities 
exceed model predictions for quasi-steady state disk evolution by more than three orders
of magnitude.  This implies that in these systems, we are actually observing the signatures
of transient events. It has been suggested that collisions of larger bodies result in episodic
dust production in debris disks 
\citep{2005ApJ...620.1010R}.  However, 
\cite{2007ApJ...658..569W} demonstrate that individual collisions are almost certainly not efficient enough to produce the observed dust overabundances, and conclude that a large-scale dynamical instability is a more likely explanation.  Furthermore, they
show that observing this phenomenon in 2\% of systems means there is a good chance
that such a cataclysmic event occurs at some point during the lifetime of {\it all} Sun-like stars.  More observations as well as modeling are required to explore this intriguing possibility.  

\acknowledgements{This work is supported by NSF Grant AST-0507727 at
Northwestern University (EWT and FAR) and by CITA (EWT).  YW is supported by a grant from NSERC.  N-body simulations were performed on CITA's McKenzie and Sunnyvale clusters.}


\end{document}